\documentclass[letter]{aa}  

\usepackage{graphicx}
\usepackage{txfonts}
\begin{document} 

   \title{Accurate spectroscopic redshift of the multiply lensed quasar PSOJ0147 from the Pan-STARRS survey}


   \author{C.-H. Lee
          \inst{1}
          }

   \institute{Subaru Telescope, National Astronomical Observatory of Japan,
              650 North Aohoku Place, Hilo, HI 96720, USA\\
              \email{leech@naoj.org}
             }

   \date{Received xxx xx, 20xx; accepted xxx xx, 20xx}

 
  \abstract
      {The gravitational lensing time delay method provides a one-step determination of the Hubble constant (H$_0$) with
        an uncertainty level \textit{on par} with the cosmic distance ladder method. However, to further investigate the nature of the dark
        energy, a H$_0$ estimate down to 1\% level is greatly needed. This requires dozens of strongly lensed quasars that
      are yet to be delivered by ongoing and forthcoming all-sky surveys.}
      {In this work we aim to determine the spectroscopic redshift of PSOJ0147, the first strongly lensed quasar candidate found in the Pan-STARRS survey. The main goal of our work
is to derive an accurate redshift estimate of the background quasar for
      cosmography.}
      {To obtain timely spectroscopically follow-up, we took advantage of the fast-track service programme that is carried out by the Nordic Optical Telescope.
        Using a grism covering 3200 - 9600 A, we identified prominent emission line features, such as Ly$\alpha$, N V, O I, C II, Si IV, C IV, and [C III] in the spectra of the background quasar of the PSOJ0147 lens system.
        This enables us to determine accurately the redshift of the background quasar.}
      {The spectrum of the background quasar exhibits prominent absorption features bluewards of the strong emission lines, such as Ly$\alpha$, N V, and C IV. These blue absorption lines indicate that the background source
        is a broad absorption line (BAL) quasar. Unfortunately, the BAL features hamper an accurate determination of redshift using the above-mentioned strong emission lines. Nevertheless, we are able to determine a redshift of 2.341$\pm$0.001 from three of the four lensed quasar images with the clean forbidden line [C III]. In addition, we also derive a maximum outflow velocity of $\sim$ 9800
km/s with the broad absorption features bluewards of the C IV emission line. This value of maximum outflow velocity is in good agreement with other BAL quasars.}


   \maketitle
%

\section{Introduction}
In the era of precision cosmology, determining H$_0$ to  1\% level locally, and comparing the local measurements
with those from the cosmic microwave background (CMB) can provide stringent constraints on the nature of dark energy, the physics of
neutrinos, and the spatial curvature of the Universe. Using a distance ladder method, the Supernovae, H0, for the Equation of State of Dark energy (SH0ES)
team has measured H$_0$ to 2.4\% \citep{2016ApJ...826...56R}, which is in tension with (at the 3.4$\sigma$ level)
the H$_0$ measurement from the Planck CMB results. While it is tempting to claim such tension is due to systematics in
the Planck measurements or a sign of new physics, it is important to obtain independent local H$_0$
measurements with accuracy and precision comparable to those of the SH0ES results.

One alternative is via gravitational lensing time delay. With this method,
for a strongly lensed, multiply imaged source, the photons from each of the images travel through different light paths (geodesics). Since the differences in the
light travel times (or time delays) only depend on the space-time curvature, we can thus
measure H$_0$. \cite{1964MNRAS.128..307R} first suggested using supernovae to measure the time delay.
However, because of the lack of all-sky transient surveys, time delay
measurements have been carried out with quasars instead.
For example, the latest study from H0LiCOW \citep{2017MNRAS.468.2590S} demonstrate that even with time delay measurements with merely three multiply lensed quasars, one can determine a H$_0$ as accurate and precise
as 3.8\% \citep{2017MNRAS.465.4914B}. This is only possible because the effects of mass along the
line of sight are also constrained with extensive spectroscopic observations \citep{2017MNRAS.467.4220R},
enabling the H0LiCOW team to carry out a detailed account of the systematics,
including the mass-sheet degeneracy.

The time delay method provides an independent and consistent local measurement of H$_0$ similar to the cosmic
distance ladder method, and results from both of these methods are in tension with the CMB results from the Planck satellite. However, we need to measure H$_0$ down
to the 1\% level to investigate the nature of the dark energy. Unfortunately, currently the H0LiCOW team only has a handful of
good multiply lensed systems, hence an increase in the number of the multiply lensed quasars is in great demand. In this regard, \cite{2017ApJ...844...90B}
have recently discovered a new quadruply lensed quasar candidate, PSOJ0147, from the Pan-STARRS survey. As three out of four lensed quasar images are rather bright, this new Pan-STARRS lensed quasar candidate provides a potentially interesting case for accurate time delay measurements.
This is especially the case in the context of the high-cadence, high S/N monitoring method recently established by \cite{2017arXiv170609424C}.
However, this is not the case for the counter image. This faintest image remains out of reach of current lens monitoring telescopes.

In this paper we present spectroscopic follow-up of this new candidate, with the aim to confirm its lensing nature, to provide an accurate spectroscopic redshift estimate, and to form a firm basis for cosmography.

\section{Observations and data reduction}
\label{sec:obs}

When \cite{2017ApJ...844...90B} first reported the discovery of PSOJ0147, it was not visible for prompt spectroscopic follow-up. The four lensed quasar images are rather bright, with brightnesses of i=15.40-16.21 mag for the three brighter images and i=17.74 for the faintest counter image. In this regard, we can easily obtain spectra with decent spectral signal-to-noise ratio (S/N) of the three brighter images. For example, the required spectral S/N can be reached with a 2 m class telescope with an exposure time of merely half an hour. 

To obtain the spectra in a timely manner, we thus made use of the fast-track service programme with the 2.5 m Nordic Optical Telescope at La Palma observatory and with its Andalucia Faint Object Spectrograph and Camera (ALFOSC).

\begin{figure}
  \centering
  \includegraphics[scale=0.4]{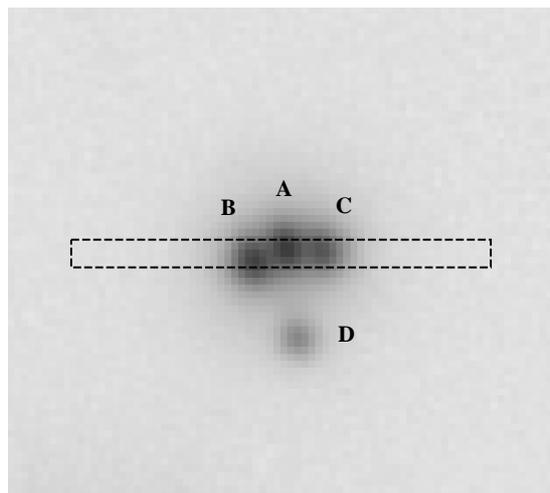}
  \caption{Set-up of our ALFOSC spectroscopic observation. The four lensed quasar images are indicated from A to D, starting from the brightest image A with i=15.40 mag. The dashed rectangle shows the 1.0 arcsec slit. With a fixed slit position angle of 90 degrees, we can obtain spectra for the three
    brighter lensed quasar images in one shot. The underlaying image is from Pan-STARRS i-band observation
  \citep{2016arXiv161205560C}, with a size of 10$\times$10 arcsec$^2$ with north up and east to the left.}
  \label{fig.slit}%
\end{figure}

To confirm the lensing nature, we only needed to observe a subset of the four lensed quasar images. We thus used a 1.0 arcsec slit to cover the three brighter lensed quasar images (see Fig. \ref{fig.slit}).
We elaborately chose a grism with a spectral range from 3200 to 9600 A. This grism enabled us to cover a wide wavelength
across the optical spectrum to identify as many quasar emission line features as possible. Given a photometric redshift estimate of z=2.6-2.8, we expected to
identify various prominent emission line features, such as Ly$\alpha$ $\lambda\lambda$ 1215.24 A,
N V $\lambda\lambda$ 1240.81 A, C IV $\lambda\lambda$ 1549.48 A, and [C III] $\lambda\lambda$ 1908.734 A.
This enables us to corroborate the Berghea et al. determination of the redshift of the quasar.
The observations were carried out on July 22, 2017 with a median seeing of 1.3 arcsec.

Since we needed to align our slit to a fixed position angle, we employed an atomspheric dispersion corrector during the spectroscopic observation to avoid differential atmospheric refraction. To get rid of 
bad pixels on the CCD and to remove cosmic ray features in the spectra, the spectroscopic observation was dithered with an A-B-B-A sequence with a dithering
size of 1 arcsec. At each dithering position we took a 300-sec exposure for each of the spectra. This resulted in a total integration time of 20 minutes.
In addition to the target, we also obtained observations of the spectroscopic standard star BD+28d4211 for flux calibration.

The data reduction was carried out in a standard manner via routines in IRAF\footnote{IRAF is distributed by the National Optical Astronomy Observatory, which is operated by the Association of Universities for Research in Astronomy (AURA) under a cooperative agreement with the National Science Foundation.}. The data reduction included steps of bias subtraction, flat fielding, wavelength calibration using a Thorium-Argon lamp, and flux calibration using the spectroscopic
standard star BD+28d4211. 

\section{Results}
\begin{figure*}[!h]
  \centering
  \includegraphics[scale=1.4]{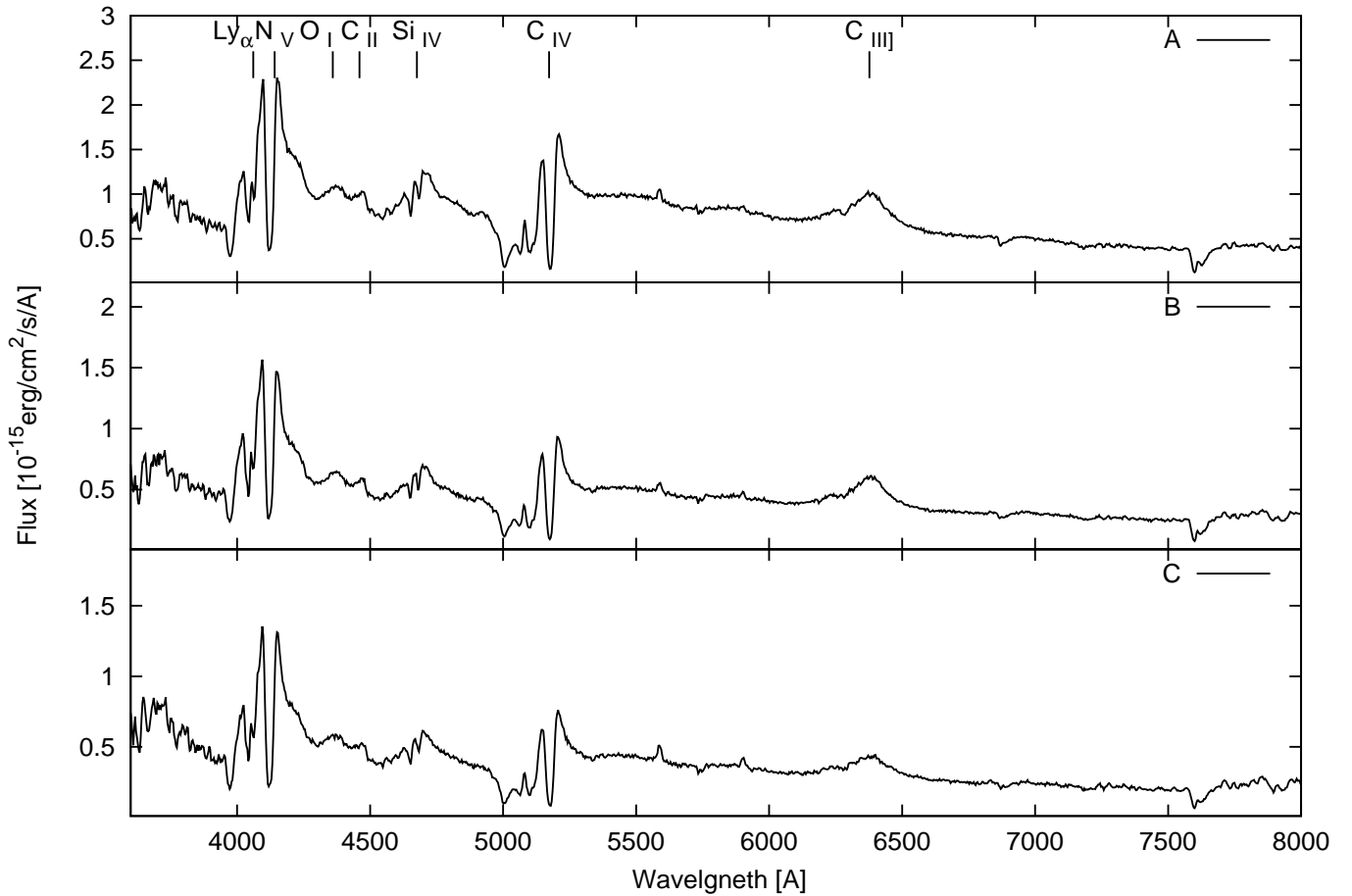}
  \caption{Spectra of images A (upper panel), B (middle panel), and C (lower panel) from ALFOSC mounted on the Nordic Optical Telescope.
    Prominent emission lines, such as Ly$\alpha$ $\lambda\lambda$ 1215.24, N V $\lambda\lambda$ 1240.81,
    Si IV $\lambda\lambda$ 1397.61, C IV $\lambda\lambda$ 1549.48, and [C III] $\lambda\lambda$ 1908.734 are indicated with black labels,
    assuming a redshift of 2.341. In addition, we also find traces of O I $\lambda\lambda$ 1305.53 and C II $\lambda\lambda$ 1335.31.
    Besides emission lines, the spectra also exhibit complicated absorption features blanketing the emission lines. This is especially the case in regions bluewards of Ly$\alpha$,
  N V, Si IV, and C IV. These broad absorption lines indicate that the background quasar of the PSOJ0147 lens system belongs to the BAL category.}
  \label{fig.spec}%
\end{figure*}

After data reduction, wavelength and flux calibration, we identify prominent emission lines, including Ly$\alpha$,
N V, Si IV, C IV, and [C III] in the spectra of all three lensed quasar images.
We also see traces of O I and C II. The spectra
from images A, B, and C, and the above-mentioned emission lines are indicated in Fig. \ref{fig.spec}.
Besides the emission lines, there are broad absorption regions bluewards
of these emission lines in the spectra of all three lensed quasar images.
These broad absorptions suggest that this quasar belongs to the broad absorption line (BAL) category.

\begin{table}
\caption{Gaussian centroids of [C III] emission line in images A-C.}
\centering
\begin{tabular}{lcc}
\hline\hline
& &  \\
Image & $\lambda_{obs}$ [A] & z\\
\hline
A & 6375.45 & 2.340\\
B & 6376.32 & 2.341\\
C & 6377.95 & 2.341\\
\hline
\hline
\end{tabular}
\tablefoot{Using the mean and standard deviation of the Gaussian
  centroids of [C III] from images A, B, and C, we derive a redshift
of 2.341$\pm$0.001.}
\label{tab.ciii}
\end{table}

The BALs originate from gaseous materials along our line of sight. Given the proximity of these BALs to the quasar emission lines in all of the spectra of the lensed quasar images, these BALs are physically associated with the
background quasar and stem from the outflow from the central AGN. 

\begin{table}
\caption{Gaussian centroids of BALs bluewards of C IV in image A-C.}
\centering
\begin{tabular}{lcc}
  \hline\hline
  & & \\
Image & $\lambda_{obs}$ [A] & Velocity [km/s] \\
\hline
A & 5008.46 &  -9756.16 \\
B & 5006.28 &  -9882.49 \\
C & 5006.55 &  -9866.84 \\
\hline
\hline
\end{tabular}
\tablefoot{Using the mean and standard deviation of the Gaussian
  centroids of C IV from images A, B, and C, we derive a maximum outflow
velocity of $\sim$ 9800 km/s.}
\label{tab.civ}
\end{table}

\label{sec:res}

Such intrinsic BALs provide important probes to the kinematics, physical conditions,
and elemental abundances in the gases surrounding AGNs,
but these BAL features hamper
an accurate redshift estimate of the background quasar.
In this regard, we thus exclude
the strong Ly$\alpha$, N V, and C IV emission lines
and only use the [C III] line for redshift estimate.
This is because [C III] is a
semi-forbidden transition, where no absorption is expected.
The [C III] emission line is therefore less susceptible to outflow
and much cleaner to serve the purpose of
an accurate redshift estimate.
The Ly$\alpha$, N V, C IV, and other fainter
emission lines can nevertheless provide a sanity check and constrain the redshift estimate from [C III].

To measure the redshift, we used the \textit{splot} task in IRAF and
fit a Gaussian function to the [C III] feature in each of the images A, B,
and C. We used the mean of the Gaussian centroids from each of the images to
determine the redshift and the standard deviation to estimate the redshift
uncertainty. We thus obtain a redshift of z=2.341$\pm$0.001. The values of the Gaussian centroids from
each of the images A, B, and C are shown in Table \ref{tab.ciii}.

Given the redshift of the background quasar, we can also estimate the
maximum outflow velocity from the BALs. Here we use the BALs bluewards of C IV
to estimate the outflow velocity. The BALs show complex feature, nevertheless, for our propose of determining the maximum outflow velocity, we only needed to use the bluest BAL feature in the vicinity of C IV. We thus only used the strong absorption line close to 5000 A at the observer's frame. We fit a Gaussian profile to each of
the spectra of images A, B, and C, and used the Gaussian centroids to obtain an maximum outflow
velocity of 9800 km/s. The values of the Gaussian centroids from
each of the images A, B, and C are shown in Table \ref{tab.civ}.

\section{Discussions}
\label{sec:dis}
An accurate redshift estimate of the background source is essential to time delay studies and pivotal to strong lens systems for cosmography.
Hence, it is important to understand the differences between our spectroscopic follow-up results and those from a previous study by \cite{2017arXiv170705873R}.
In \cite{2017arXiv170705873R}, they have obtained a higher redshift estimate. Here we list two possible causes for such discrepancies in the
redshift estimate from \cite{2017arXiv170705873R}:

\begin{enumerate}
\item \textbf{Narrow wavelength coverage.} While the Keck Cosmic Web Imager provides a spectral resolution of R$\sim$5000,
  the spectral coverage in \cite{2017arXiv170705873R} is rather narrow,  from 3500 to 5500 A. In this regard, \cite{2017arXiv170705873R}
  have only identified Ly$\alpha$, Si IV, and C IV in their spectra and have also pointed out that observations at longer wavelengths
  are needed to better constrain the redshift of the background quasar.
  Our spectra, on the other hand, cover a wider spectral range from 3200 to
  9600 A. Our spectra are thus suffice for a more accurate spectroscopic redshift estimate. \\

\item \textbf{BAL contaminations.} As has been pointed out by \cite{2017arXiv170705873R}, the background quasar of PSOJ0147 belongs
  to the BAL category, showing complicated absorption features blanketing the broad emission lines of Ly$\alpha$, Si IV, and
  C IV. In this regard, BALs hamper accurate redshift estimates if these strong emission lines are used for spectroscopic
  redshift determination. While \cite{2017arXiv170705873R} have excluded spectral range bluer than $\lambda_{rest}$=1250 A, i.e. the
  Ly$\alpha$ and N V regions, they have nevertheless included C IV for redshift estimate. The C IV emission line is the brightest after Ly$\alpha$ and N V, and is still
  heavily blanketed by BALs, thus hampering accurate redshift determination as well. Our redshift estimate, on the other hand,
  follows the approach of \cite{2017ApJ...838L..15L} and only relies on the [C III] emission line. As [C III] is a semi-forbidden line,
  it is less susceptible to BALs and is one of the cleanest emission lines for redshift determination. 
  \end{enumerate}

As a comparison, we also performed a test of the redshift estimate by adopting the same spectral range to our spectra as that of Rubin et al. (2017), and we used this
  narrower spectral range to investigate the redshift estimate from our data. In this spectral range, the strongest emission line feature is the C IV emission line. We thus calculated a redshift with the peak of the C IV emission line in our data, which is at 5215 A in the observer's frame. We then obtained a redshift estimate of z=2.37, which is in agreement with the redshift estimate from Rubin et al. (2017). This suggests that the discrepancy between our redshift measurment and that of Rubin et al. (2017) is actually due to the spectral range.

In addition to the strong lensing effect, the background quasar may also be
subject to microlensing effects. As microlensing has different effects
on the continuum than on the emission lines and absorption lines, one can thus use the ratio spectra to unveil the microlensing effects. However, owing to the poor seeing condition and blending of our spectra, we were not able to further investigate the microlensing effects with the current data.

\section{Conclusions}
\label{sec:con}

PSOJ0147 is a recently discovered and first quadruply lensed quasar candidate from the Pan-STARRS survey public data.
In this work, we obtained timely spectroscopic follow-up of PSOJ0147 using the ALFOSC on board the Nordic Optical Telescope.
Our results can be summarized as follows:
\begin{enumerate}

\item
  \textit{We provide a more accurate redshift estimate of the background lensed quasar.} Given the BAL nature of the background quasar,
  it is difficult to determine the redshift of the background quasar using, e.g. Ly$\alpha$, N V, Si IV, and C IV emission
  lines. Unfortunately, the \cite{2017arXiv170705873R} study has only covered a narrow wavelength (i.e. up to C IV at the observer's
  frame), which hampered an accurate redshift determination. In this work, we present a much wider spectral coverage. In addition, our redshift estimate is only
  based on the semi-forbidden emission line [C III], which is less susceptible and much cleaner for the purpose of redshift determination.
  We estimate the redshift of the background quasar to be 2.341$\pm$0.001, providing a firm base to use the PSOJ0147 lens system for
  cosmography studies. \\
  
\item
  \textit{We determine the maximum outflow velocity of the background quasar.} The BAL quasars are instrumental in probing gases around AGNs, especially the gaseous
  kinematics, physical conditions, and elemental abundances. Using the BALs bluewards of the C IV emission line, we are able
  to derive an maximum outflow velocity of $\sim$ 9800 km/s for the background quasar of PSOJ0147. Further studies of the surrounding gases
  will require spectroscopic observations with higher resolutions. \\

\end{enumerate}

The fact that the background quasar of PSOJ0147 belongs to the BAL category is an important discovery, especially in two aspects.
  First of all, BAL quasars exhibit more variability than other quasars, hence BAL quasars are ideal targets for time delay measurements. Secondly, microlensing effects in the BAL part of the quasar spectra can provide insights into the BAL structure.

Besides accurate redshift, time delay measurements are another essential ingredient in the use of multiply lensed quasars for cosmography. While
previous works rely on decade-long photometric monitoring campaigns to obtain precise time delay measurements, it has been shown that
high-cadence, high S/N photometric follow-up can drastically reduce the required time span of photometric monitoring down to a few months
\citep{2017arXiv170609424C}. With this high-cadence approach, we can even beat down the microlensing effect, further reducing the systematics
in time delay analyses. The PSOJ0147 lens system is suitable for such a short time-span approach thanks to its bright nature. With a magnitude of 15.4-16.2
in the i-band for the three brighter images, we can reach S/N $>$ 1000 within a few minutes using 2 m class telescopes, making it another ideal lens system
for gravitational lensing time delay and cosmography. The faintest counter image, however, remains out of reach of current lens monitoring telescopes.

\begin{acknowledgements}
  We are indebted to the referee, whose insightful comments greatly improved
  this manuscript. Based on observations made with the Nordic Optical Telescope, operated by the Nordic Optical Telescope Scientific Association at the Observatorio del Roque de los Muchachos, La Palma, Spain, of the Instituto de Astrofisica de Canarias. The data presented here were obtained with ALFOSC, which is provided by the Instituto de Astrofisica de Andalucia (IAA) under a joint agreement with the University of Copenhagen and NOTSA.
\end{acknowledgements}

%
%

\end{document}